\documentclass[twocolumn,showpacs,preprintnumbers,amsmath,amssymb]{revtex4}

\usepackage{graphicx}
\usepackage{dcolumn}
\usepackage{bm}

\begin{document}

\title{Origin of the Scaling Law in Human Mobility: Hierarchical Organization of Traffic Systems}

\author{Xiao-Pu Han$^{1}$}
\author{Qiang Hao$^{1}$}
\author{Bing-Hong Wang$^{1,2}$}
\author{Tao Zhou$^{1,3}$}

\affiliation{$^{1}$Department of Modern Physics, University of
Science and Technology of China, Hefei 230026, People's Republic of
China\\$^{2}$Research Center for Complex System Science, University
of Shanghai for Science and Technology, Shanghai 200093, People's
Republic of China \\$^{3}$Department of Physics, University of
Fribourg, Fribourg 1700, Switzerland}

\date{\today}

\begin{abstract}
Uncovering the mechanism leading to the scaling law in human
trajectories is of fundamental importance in understanding many
spatiotemporal phenomena. We propose a hierarchical geographical
model to mimic the real traffic system, upon which a random walker
will generate a power-law travel displacement distribution with
exponent -2. When considering the inhomogeneities of cities'
locations and attractions, this model reproduces a power-law
displacement distribution with an exponential cutoff, as well as a
scaling behavior in the probability density of having traveled a
certain distance at a certain time. Our results agree very well with
the empirical observations reported in [D. Brockmann \emph{et al.},
Nature \textbf{439}, 462 (2006)].
\end{abstract}

\pacs{89.75.Fb, 05.40.Fb, 89.75.Da}

\maketitle

Studies on the non-Poisson statistics of human behaviors have
recently attracted much attention \cite{bara,vqz1,zhou}. Besides the
inter-event or waiting time distribution, the spatial movements of
human also exhibit non-Poisson statistics. Brockmann \emph{et al.}
\cite{bro} investigated the bank note dispersal, as a proxy for
human movements, and revealed indirectly a power-law distribution of
human travel displacements. Gonzalez \emph{et al.} \cite{gon}
studied the human travel patterns by measuring the distance of
mobile phone users' movements in different stations, and observed a
similar scaling law. Actually, the mobility patterns of many animals
also show power-law-like displacement distributions
\cite{bart1,ramo,sims}. The ubiquity of such kind of distributions
attracts scientists to dig into the underlying mechanism. Some
interpretations, such as optimal search strategy \cite{visw,bart2},
olfactory-driven foraging \cite{reyn} and deterministic walks
\cite{santo}, have already been raised for the power-law
displacement distribution in animals mobility patterns, however,
they are based on the prey processes and thus cannot be used to
explain the observed scaling law in human trajectories, which is
still an open problem. In this paper, we propose a model to mimic
the human travel pattern, where the hierarchical organization of the
real human traffic systems is taken into account. Our model can
reproduce the power-law displacement distributions, as well as the
scaling behavior in probability density of having traveled a certain
distance at a certain time, agreeing very well with the empirical
results reported in Ref. \cite{bro}.

\begin{figure}
\includegraphics[width=8.0cm]{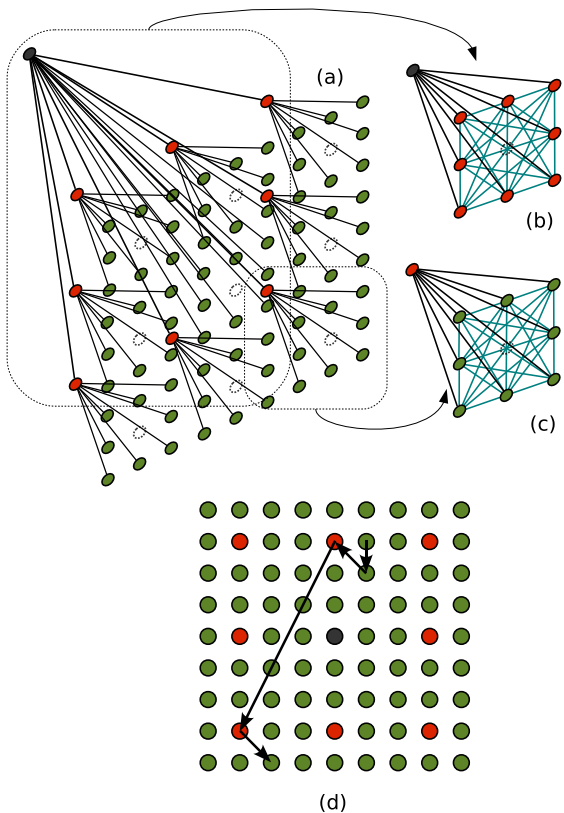}\\
\caption{(Color online) Illustration of the hierarchical structure
and moving rules in the present model. (a)-(c) show the connections
among the different-layer cities (black lines) and the same-layer
cities (blue lines). Where the green, red, and black circles denote
the cities in 3rd, 2nd and 1st layers, respectively. (d) gives a
typical trajectory of a walker moving from a lower-layer city to a
higher-layer city in other sub-region. A fractal network model with
similar motivation can be found in Ref. \cite{Kalapala2006}.}
\end{figure}

Let's think about the real human traffic systems. Generally
speaking, a district (e.g., a province or a state) usually has a
core city, like its capital; around this core city, there are
several big cities as the secondary centers (e.g., municipalities);
then, each of these centers is rounded by some counties; and towns
and villages will surround each of the counties. A hierarchical
traffic system is built accordingly. Imaging people traveling from a
town, $a$, subordinating to the central city, $A$, to another town
$b$ that is subordinated to the central city $B$. There is usually
no direct way connecting $a$ and $b$, and the typical route is
$a\rightarrow A\rightarrow B\rightarrow b$. This kind of
hierarchical organization is not just inside a country or a
district, but across the whole world. For example, if one wants to
travel from the \emph{University of Science and Technology of China}
to the \emph{University of Fribourg}, there is no direct way
connecting \emph{Hefei} and \emph{Fribourg}, instead, one has to
follow the route
\emph{Hefei}$\rightarrow$\emph{Shanghai}$\rightarrow$\emph{Z\"urich}$\rightarrow$\emph{Fribourg}
although it is much longer than the geographical distance between
\emph{Hefei} and \emph{Fribourg}. Such a hierarchical organization
as well as the resulting scale invariance in road networks have
already been demonstrated recently \cite{Kalapala2006}.

For simplicity, we call all the units \emph{cities}. In our model,
cities are organized in $N$ layers. A uniform 3-layer system is
shown in Fig. 1, in which, 81 cities locate on the centers of a $9
\times 9$ lattice. The most central city is put in the first layer,
and the whole region is divided into 9 sub-regions, each contains a
$3 \times 3$ lattice. Except the middle sub-region, all other eight
sub-regions have their own central cities, namely the second layer
cities, which locate at the centers of those sub-regions. Meanwhile,
there are eight third layer cities around each of the second layer
cities as well as the first layer city. An illustration is shown in
Fig. 1. Denote $N$ the number of layers and $K$ the number of first
layer cities. We assign $M$ sub-regions to each of the $K$ 1st-layer
cities, with the 1st-layer city locating in the center, and $(M-1)$
2nd-layer cities are respectively put in the remain $(M-1)$
sub-regions. Each of the $KM$ sub-regions is further divided into
$M$ sub-sub-regions, with the 1st- or 2nd-layer cities locating in
the center and $(M-1)$ newly generated 3rd-layer cities put in the
remain $(M-1)$ sub-sub-regions. Repeating this process until the
$N$th-layer cities are generated. For $2\leq n \leq N$, there are
$KM^{n-2}(M-1)$ $n$th-layer cities. Note that, in this model, to
make sure the lattice is fulfilled with cities, $M$ must be equal to
$(2q-1)^2$ where $q$ is a certain integer larger than 1.

The $K$ first layer cities are fully connected with each other, each
of which is connected with the nearest $(M-1)$ $N$th-layer cities
and the nearest $(M-1)$ second layer cities. Each of the second
layer cities is connected with the nearest $(M-1)$ $N$th-layer
cities, the nearest $(M-1)$ third layer cities, as well as the other
$(M-2)$ second layer cities belonging to the same first layer city.
Actually, for $1\leq n < N$, each of the $n$th-layer cities is
connected with the nearest $(M-1)$ $N$th-layer cities, the nearest
$(M-1)$ $(n+1)$th-layer cities, as well as the other $(M-2)$
$n$th-layer cities belong to the same $(n-1)$th-layer city. Note
that, all the connections are symmetry, and the direct move between
two cities is allowed only if they are connected. Figure 1
illustrates an example with $N=3$, $K=1$ and $M=9$. The modeled
system can be viewed as a hierarchical network. Different from the
real hierarchical networks \cite{rav2} or the mathematical models
\cite{an,ran}, it is hierarchical but not scale-free.

\begin{figure}
\includegraphics[width=8.6cm]{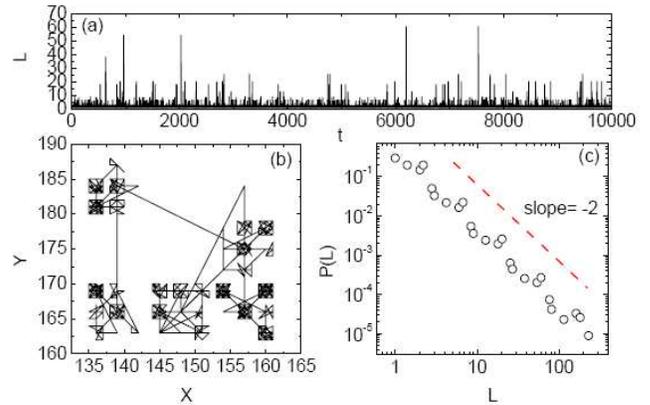}\\
\caption{(Color online) Burstiness of long-range travels. Parameters
are set as $N = 5$, $M = 9$, and $K = 9$. We choose $N=5$ because it
is typical in the real world, such as
province-city-county-town-village in China and
region-department-arrondissement-canton-commune in France. (a) The
sequence of the displacement, $L$, of a random walker in $10^4$
consecutive steps. (b) The walker's trajectory in 2000 consecutive
steps. (c) The distribution $P(L)$, where the dash line, as a guide
for eyes, is of slope -2. The data in panel (c) is obtained by
averaging over 100 independent runs, and each lasts $10^5$ steps.}
\end{figure}

We consider the simplest case where a random walker is consequently
moving from one city to a random neighboring city (two cities are
said to be neighboring if they are connected). Figure 1(d) shows a
typical trajectory of a walker moving from a lower-layer city to a
higher-layer city in other sub-region. The displacement of the
walker in one step is defined as the geometric distance $L$, and the
distribution of $L$ is what we mainly concern in this paper. As
shown in Fig. 2, burstiness of long-range travels is clearly
observed and the distribution of travel displacement, $P(L)$,
approximately obeys a power-law form with exponent -2.

The essential physics of this model is a random-walk process in a
geographical network where edges are of different geometric
distances. For a random walker in a connected symmetry (undirected)
network, in the long time limit, each edge has the same chance to be
visited (this proposition is hold even for a very heterogenous
network, since for an arbitrary node, the number of times being
visited is proportional to its degree while the probability that a
specific adjacent edge of this node is consecutively visited is
inversely proportional to the degree. Details can be found in Ref.
\cite{Lovasz1996}). Therefore, the displacement distribution of a
random walker is equivalent to the distribution of edges' geometric
distances. Let $d_n$ $(n>1)$ denote the average geometric distance
of edges connecting two $n$th-layer cities (they must belong to the
same $(n-1)$th-layer city) and an $n$th-layer city and an
$(n-1)$th-layer city (the former belongs to the latter), and $m_n$
denote the total number of edges contributed to $d_n$. Obviously,
for $1<n\leq N$, $d_n=d_N\sqrt{M}^{N-n}$. $M$, $N$ and $d_N$ can be
considered as constants, and thus we have $d_n\sim M^{-n/2}$. On the
other hand, for $n>1$, $m_{n} = KC_{M}^{2}M^{n-2}$ where
$C_M^2=M(M-1)/2$. That is, $m_n\sim M^n$. Roughly speaking, $d_n$
and $m_n$ play the roles of geometric distance and the number of
edges associated with such a distance. As we have already obtained
the scaling $m_n\sim d_n^{-2}$, we can deductive that the
displacement distribution for a system with sufficiently large $N$
and in the long time limit obeys the scaling $P(L)\sim L^{-2}$,
which is in accordance with the simulation result shown in Fig.
2(c). The observed result implies that the scaling law in human
trajectories may results from the inherent hierarchical organization
in traffic systems.

\begin{figure}
\includegraphics[width=8.6cm]{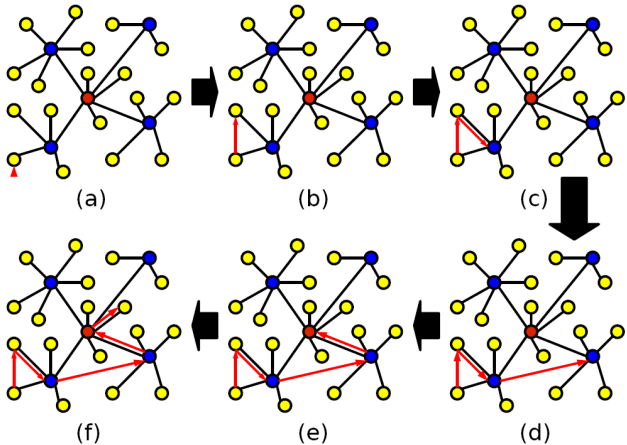}\\
\caption{(Color online) Illustration of a typical five-step
movements in a 3-layer inhomogeneous model, where red, blue, and
yellow respectively denote the 1st-layer, 2nd-layer and 3rd-layer
cities. For clarity, we do not plot the connections between cities
in the same layer.}
\end{figure}

\begin{figure}
\includegraphics[width=8.6cm]{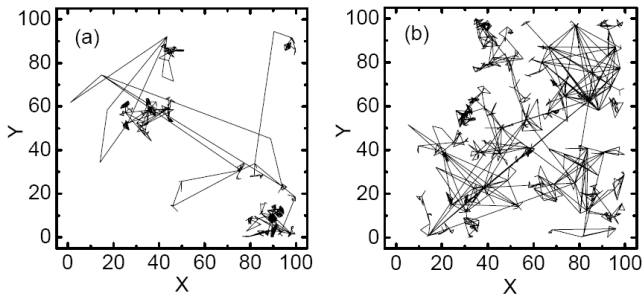}\\
\caption{The trajectory of a walker in 2000 consecutive steps in the
heterogeneous model with parameter setting $N = 5$, $M = 9$, $K =
9$, and $S = 100$. (a) and (b) respectively correspond to the cases
of $r = 1$ and $r = 2$.}
\end{figure}

\begin{figure}
\includegraphics[width=9cm]{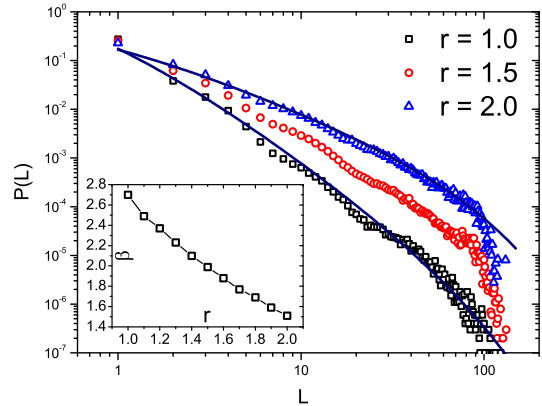}\\
\caption{(Color online) Displacement distribution, $P(L)$, for
inhomogeneous model, obtained by 100 independent runs, each of which
lasts $10^5$ time steps. The parameters are $N=5$, $M=9$, $K=9$ and
$S=100$. Red circles, green triangles and violet rectangles
correspond to the cases of $r=2.0$, $r=1.5$, and $r=1.0$,
respectively. The blue lines denote the fitting functions, $f(L) =
0.32(L + 0.52)^{-1.51}e^{-L/60}$ for $r = 2.0$, and $f(L) = 0.53(L +
0.50)^{-2.70}e^{-L/55}$ for $r = 1.0$. The inset shows the relation
between the power-law exponent $\beta$ and the parameter $r$.}
\end{figure}

Although the present model can reproduce the power-law displacement
distribution, the absolute value of exponent is higher than the
empirical ones \cite{bro,gon}, and the model is obviously
oversimplified. Firstly, real cities are not located in a completely
uniform matter, but with some irregularity. Secondly, the model
assumes that each city has the same attraction for the walker,
however, in the real world, a central city is generally much more
attractive than a small town. We next propose a modified model
taking into consideration the inhomogeneous locations and
attractions of cities. In this \emph{inhomogeneous model}, all
cities are randomly distributed in an $S\times S$ square (we keep
the number of cities in each layer the same as the original model),
each of the $n$th-layer cities $(2\leq n \leq N)$ is connected to
the nearest higher-layer city, and two $n$th-layer cities are
connected if they are connected to the same higher-layer city. All
the $K$ 1st-layer cities are fully connected to each other.

As mentioned above, the center city should have greater attraction,
which is represented by a layer-dependent weight, $w_n=r^{N-n}$,
where $n$ denotes the layer and $r\geq 1$ is a free parameter. The
probability that the walker will move along an edge is proportional
to its weight (A similar weighted random walk model has previously
been proposed to explain the nonlinear dependence of the airport
throughput on the connectivity \cite{Ou2007}). Clearly, the larger
$r$ indicates higher heterogeneity. In Fig. 3, we show an
illustration of a typical trajectory in a 3-layer inhomogeneous
model, and in Fig. 4 we report the trajectories for $r=1$ and $r=2$
in a 5-layer inhomogeneous model. As shown in Fig. 5, for the
inhomogeneous model, the displacement distribution, $P(L)$, is still
heavy-tailed and can be well fitted by a power-law function with an
exponential cutoff, as $P(L)=c(a+L)^{-\beta}\texttt{e}^{-x/x_c}$. In
addition, as shown in the inset of Fig. 5, when $r$ increases from
1.0 to 2.0, the power-law exponent, $\beta$, monotonously decreases
from 2.70 to 1.51, covering the range of empirical observations
\cite{bro,gon}. This result suggests that the inhomogeneity in
cities' attractions may be the reason why the absolute value of
power-law exponent in the real human displacement distribution is
lower than that predicted by the homogeneous model (i.e., 2.0),
while the inhomogeneity of cities' locations enlarges the absolute
value of such exponent.

Finally, we check whether our model can reproduce the spatiotemporal
statistics of real human mobility. Providing the trajectory of a
random walker, one can obtain the probability $W(d,t)$ of having
traveled a distance $d$ at time $t$ (the same technique has been
adopted in preparing Fig. 2a in Ref. \cite{bro}, please see details
there). As shown in Fig. 6, a scaling behavior $r(t)\sim t^\alpha$
with $\alpha \approx 1.0$ is clearly observed, which agrees well
with the empirical result, $\alpha\approx 0.95$, reported in Ref.
\cite{bro}. Similar scaling behavior can also be observed for
$r=1.0$, however, the exponent, $\alpha \approx 0.5$, is far less
than the empirical value. In addition, providing the travel
displacement distribution, this scaling behavior with $\alpha$
around 1.0 can not be reproduced by a L\'evy flight \cite{bro}.

\begin{figure}
\includegraphics[width=9cm]{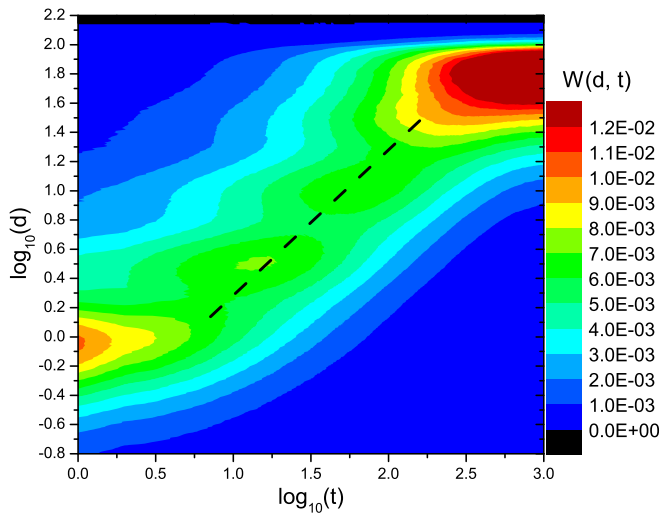}\\
\caption{(Color online) The probability $W(d,t)$ of having traveled
a distance $d$ at time $t$. The parameters are $N=5$, $M=9$, $K=9$,
$S=100$ and $r=2.0$. This plot is obtained by averaging 1000
independent runs, each of which lasts $10^4$ time steps. The black
dash line, as a guide of eyes, is of slope 1.}
\end{figure}

Uncovering the human traveling pattern is of fundamental importance
in the understanding of various spatiotemporal phenomena
\cite{bro,gon}, and may finds applications in the design of traffic
systems \cite{Barthelemy2006}, the control of human infectious
disease \cite{Hufnagel2004}, the military service planning
\cite{Zhao2008}, and so on. Although empirical results about the
scaling law of long-range human travels have been reported for
years, it lacks the understanding of the underlying mechanism. This
work gives raise to a very possible reason causing the heavy-tailed
displacement distribution, $P(L)$, that is, the hierarchical
organization of traffic systems. The secondary ingredient, also
playing appreciable role in determining the traveling pattern, is
the inhomogeneities of the locations and attractions of cities: The
former enlarges the exponent $\beta$, while the latter depresses it
(essentially, the inhomogeneity of attractions results from the
inhomogeneous population density and economic development).
Actually, as shown in Fig. 5, with tunable strength of
inhomogeneity, the exponent $\beta$ is also tunable. When $r=2$,
meaning the topper-layer cities having greater attractions, the
statistical features produced by our model are very close to the
empirical ones reported in Ref. \cite{bro}, not only the
displacement distribution, but also the spatiotemporal statistics of
mobility.

We acknowledge Changsong Zhou and Aaron Clauset for valuable
suggestions. This work is supported by 973 program (2006CB705500),
and the National Natural Science Foundation of China (10532060,
10635040 and 70871082). Authors are ordered in alphabet.

\end{document}